\documentclass{article}

\usepackage[dvipsnames]{xcolor}
 
\usepackage{tikz-cd} 

\usepackage{amsmath,amssymb,amsfonts} % Math symbols and environments
\newcommand{\cset}[2]{\left\{ #1 \; \middle| \; #2 \right\}}

\newcommand{\SE}{\mathbf{SE}}
\newcommand{\SIM}{\mathbf{SIM}}
\usepackage{arxiv}

\usepackage[utf8]{inputenc} % allow utf-8 input
\usepackage[T1]{fontenc}    % use 8-bit T1 fonts
\usepackage{hyperref}       % hyperlinks
\usepackage{url}            % simple URL typesetting
\usepackage{booktabs}       % professional-quality tables
\usepackage{amsfonts}       % blackboard math symbols
\usepackage{nicefrac}       % compact symbols for 1/2, etc.
\usepackage{microtype}      % microtypography
\usepackage{lipsum}		% Can be removed after putting your text content
\usepackage{graphicx}
\usepackage[numbers]{natbib}
\usepackage{doi}
\usepackage{amsthm}
\newtheorem{thm}{Theorem}

\title{Synchronous Observer Design for Landmark-Inertial SLAM with Almost-Global Convergence \thanks{This work has been submitted to IFAC for possible publication. This research was supported by the Horizon Europe MSCA PF MEW (101154194).
Arkadeep Saha was partially supported by the Institute of Eminence
Funding of IIT-Bombay for his stay and visit to University of Twente.} }

\author{ \href{https://orcid.org/0009-0000-7524-3072}{\includegraphics[scale=0.06]{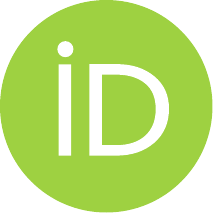}\hspace{1mm}Arkadeep~Saha}\\
	Centre for Systems and Control\\
	Indian Institute of Technology Bombay\\
	Mumbai-400076, India \\
	\texttt{22b1270@iitb.ac.in} \\
	\And
	\href{https://orcid.org/0000-0003-4391-7014}{\includegraphics[scale=0.06]{orcid.pdf}\hspace{1mm}Pieter van~Goor} \\
	Robotics and Mechatronics (RaM) group\\
	EEMCS Faculty, 
	University of Twente\\
    Enschede, The Netherlands\\
	\texttt{p.c.h.vangoor@utwente.nl} \\
    \And
	\href{https://orcid.org/0000-0002-5670-1282}{\includegraphics[scale=0.06]{orcid.pdf}\hspace{1mm}Antonio~Franchi} \\
    Robotics and Mechatronics (RaM) Group, \\EEMCS Faculty, University of Twente, \\Enschede, The Netherlands\\ \texttt{a.franchi@utwente.nl}\\
{\small and}\\
DIAG, Sapienza University of Rome, \\ 00185 Rome, Italy\\
\texttt{antonio.franchi@uniroma1.it}
    \And
    \href{https://orcid.org/0000-0002-5746-7096}{\includegraphics[scale=0.06]{orcid.pdf}\hspace{1mm}Ravi~Banavar}\\
	Centre for Systems and Control\\
	Indian Institute of Technology Bombay\\
	Mumbai-400076, India \\
	\texttt{banavar@iitb.ac.in} \\
}

\renewcommand{\shorttitle}{}

\hypersetup{
pdftitle={A template for the arxiv style},
pdfsubject={q-bio.NC, q-bio.QM},
pdfauthor={David S.~Hippocampus, Elias D.~Striatum},
pdfkeywords={First keyword, Second keyword, More},
}

\begin{document}
\maketitle

\begin{abstract}
	Landmark Inertial Simultaneous Localisation and Mapping (LI-SLAM) is the problem of estimating the locations of landmarks in the environment and the robot's pose relative to those landmarks using landmark position measurements and measurements from Inertial Measurement Unit (IMU). This paper proposes a nonlinear observer for LI-SLAM posed in continuous time and analyses the observer in a base space that encodes all the observable states of LI-SLAM. The local exponential stability and almost-global asymptotic stability of the error dynamics in base space is established in the proof section and validated using simulations. 
\end{abstract}

% keywords can be removed

\section{Introduction}
Simultaneous Localisation and Mapping (SLAM) is the problem of estimating an environment map while concurrently estimating a robot's pose with respect to this map, and has been an active area of research in mobile robotics since last thirty years \citep{bailey2006slam:part1}.
The two main approaches to SLAM are the extended Kalman filter (EKF) \citep{bailey2006slam:part1} and graph-based nonlinear optimization \citep{kaess2012isam2}, both of which have different advantages and drawbacks.
The EKF-based SLAM approaches traditionally suffer from statistical inconsistency \citep{huang2007convergence,lee2006observability}, while the optimization-based approaches 
instead exhibit high computational complexity and limited robustness \citep{cadena2017past}. 
This has driven recent interest from the nonlinear observer community, where geometric methods have provided new solutions to SLAM with guarantees of stability and consistency \citep{barrau2015ekf,johansen2016globally,mahony2017geometric,zlotnik2018gradient}.

Landmark-Inertial SLAM (LI-SLAM) is a version of the SLAM problem where the available measurements include angular velocity and acceleration from an Inertial Measurement Unit (IMU) alongside complementary exteroceptive landmark position measurements, such as those that may be provided by a stereo camera, an RGB-D camera, or a lidar.
This problem has been studied by the nonlinear observer community using a variety of Kalman filters.
\cite{lourencco20133} proposed a globally asymptotically stable (GAS) Kalman filter for LI-SLAM expressed in the body-frame of the robot, and formulated a Procrustes problem to estimate the robot's pose with respect to the inertial frame.
\cite{johansen2016globally} addressed LI-SLAM by considering additionally a magnetometer which they used in an attitude heading reference system (AHRS) to estimate the robot's attitude, after which they solved the remaining mapping and position estimation problem using linear time-varying Kalman filter.
\cite{barrau2015ekf} applied the invariant EKF (IEKF) for landmark SLAM (where the robot's body-frame velocity measurement is assumed to be available) by introducing a novel Lie group $\mathbf{SE}_{n+1}(3)$ and showing that this led to desirable group-affine dynamics.
Each of these approaches face limitations with all of them requiring at least a quadratically scaling computational complexity associated with the Kalman filter.

Deterministic nonlinear observers for SLAM follow from a rich history of geometric observers for attitude estimation \citep{bonnabel2008symmetry, mahony2008nonlinear} and pose estimation \citep{baldwin2009nonlinear, vasconcelos2010nonlinear} using Lie groups.
\cite{mahony2017geometric} designed an observer for kinematic landmark SLAM by introducing the $\mathbf{SLAM}_n(3)$ Lie group and defining a quotient manifold structure to encode the invariance of SLAM to changes in the inertial reference frame that led to inconsistency issues in classic approaches.
In closely related work, \cite{zlotnik2018gradient} developed a gradient-based observer for SLAM that also includes estimation of biases in linear and angular velocity inputs. 
\cite{wang2018geometric} build on both of these works by designing their observer on the matrix Lie group $\mathbf{SE}_{n+1}(3)$ and by considering landmarks with input velocities.
\cite{JoshiBundle} provided a fibre bundle framework for analysing the SLAM problem with kinematic landmarks. 
\cite{boughellaba2025nonlinear} has recently proposed a nonlinear observer on the $\mathbf{SE}_{n+3}(3)$ Lie group for Landmark-inertial SLAM with almost-global asymptotic stability, which uses gravity direction as an additional auxiliary state in the observer.

In this paper, we present a nonlinear geometric observer for the landmark-inertial SLAM (LI-SLAM) problem posed in continuous time for static environments. 
The approach extends the synchronous observer design for Inertial Navigation Systems presented in \citep{VANGOOR2025112328} to also include the positions of the landmarks in the state.
Based on the SLAM manifold introduced in \citep{mahony2017geometric}, a quotient manifold structure is developed to represent the LI-SLAM state-space.
A novel manifold, termed the \emph{LI-SLAM base space} $\mathcal{M}_n^{LI}(3)$ is introduced to uniquely identify all observable states of the LI-SLAM problem, using a projection map from the total space to the base space.
The base space encodes the natural invariance of the LI-SLAM problem under rotation about the vertical axis and translation of the inertial frame.
The resulting observer is shown to have an almost globally asymptotically stable and locally exponentially stable error system in the base space. 

This paper consists of four sections alongside the introduction and the conclusion. 
Section \ref{sec:preliminaries} introduces the mathematical preliminaries and notations used in the paper. 
Section \ref{sec:problem_description} provides the description, invariance and Lie group interpretation of the LI-SLAM problem. 
In Section \ref{sec:observer_design}, we provide the observer design and the proofs of stability and convergence. 
The simulation results are provided in Section \ref{sec:simulations}, verifying the theory developed throughout the paper.

\section{Preliminaries}\label{sec:preliminaries}
The special orthogonal group is the Lie group of 3D rotations,
\begin{align*}
    \mathbf{SO}(3):= \{R \in \mathbb{R}^{3 \times 3}|\ R^\top R = I_3,\ \text{det}(R)=1 \}.
\end{align*}
For any vector $\Omega\in \mathbb{R}^3$, define
\begin{align*}
    \Omega^\times = \begin{pmatrix}
        0 & -\Omega_3 & \Omega_2\\
        \Omega_3 & 0 & -\Omega_1\\
        -\Omega_2 & \Omega_1 & 0
    \end{pmatrix}.
\end{align*}
Then $\Omega^\times v = \Omega\times v$ for any $v\in \mathbb{R}^3$ where $\times$ is the usual vector(cross) product. The Lie algebra of $\mathbf{SO}(3)$ is defined
\begin{align*}
    \mathfrak{so}(3):= \{\Omega^\times \in \mathbb{R}^{3\times3}\ |\ \Omega\in \mathbb{R}^3\}.
\end{align*}
For any two vectors $a,b \in \mathbb{R}^3$, one has the following identities:
\begin{align*}
    a^\times b= -b^\times a, && (a^\times)^\top=-a^\times, \\
    a^\times b^\times=ba^\top-a^\top bI_3, && (a\times b)^\times = ba^\top-ab^\top.
\end{align*}
The extended special Euclidean group and its Lie algebra are defined
\begin{align*}
    \mathbf{SE}_n(3) &:=\left\{ \begin{pmatrix}
        R & V\\
        0_{n\times3} & I_n
    \end{pmatrix} \middle|\ R \in \mathbf{SO}(3),\ V\in \mathbb{R}^{3\times n} \right\},\\
    \mathfrak{se}_n(3) &:=\left\{ \begin{pmatrix}
        \Omega^\times & W\\
        0_{n\times3} & 0_{n\times n}
    \end{pmatrix} \middle|\ \Omega \in \mathbb{R}^3,\ W\in \mathbb{R}^{3\times n} \right\}.
\end{align*}

An element of $\mathbf{SE}_n(3)$ may be denoted $X = (R, V)$ for convenience, where $R \in \mathbf{SO}(3)$ and $V\in \mathbb{R}^{3\times n}$. Likewise, an element of $\mathfrak{se}_n(3)$ can be denoted by $\Delta = (\Omega_\Delta, W_\Delta)$, where $\Omega_\Delta \in \mathbb{R}^3$ and $W_\Delta \in \mathbb{R}^{3\times n}$. The matrix Lie group $\mathbf{SIM}_n(3)$ and its Lie algebra $\mathfrak{sim}_n(3)$ are defined by \citep{van2023constructiveINS} 
\begin{align*}
    \mathbf{SIM}_n(3) &:=  \left\{ \begin{pmatrix}
        R & V\\
        0_{n\times3} & A
    \end{pmatrix} \Bigg|\ R \in \mathbf{SO}(3),\ V\in \mathbb{R}^{3\times n}, A\in \mathbf{GL}(n) \right\},\\
    \mathfrak{sim}_n(3) &:= \left\{ \begin{pmatrix}
        \Omega^\times & W\\
        0_{n\times3} & S
    \end{pmatrix} \Bigg|\ \Omega \in \mathbb{R}^3,\ W\in \mathbb{R}^{3\times n}, \ S \in \mathfrak{gl}(n) \right\}.
\end{align*}
An element of $\mathbf{SIM}_n(3)$ can be denoted $Z = (R_Z, V_Z, A_Z)$ for convenience, where $R_Z\in \mathbf{SO}(3)$, $V_Z\in \mathbb{R}^{3\times n}$ and $A_Z \in \mathbf{GL}(n)$. Likewise, an element of $\mathfrak{se}_n(3)$ can be denoted by $\Gamma = (\Omega_\Gamma, W_\Gamma, S_\Gamma)$, where $\Omega_\Gamma \in \mathbb{R}^3$, $W_\Gamma \in \mathbb{R}^{3\times n}$ and $S_\Gamma \in \mathfrak{gl}(n)$. Let $\sigma_Z:\mathbf{SE}_n(3)\rightarrow \mathbf{SE}_n(3)$ be defined by $\sigma_Z(X):= ZXZ^{-1}$, in the sense of matrix multiplication, where $Z\in \mathbf{SIM}_n(3)$. 
\\The automorphism of $\mathbf{SE}_n(3)$ is a diffeomorphism $\sigma:\mathbf{SE}_n(3)\rightarrow \mathbf{SE}_n(3)$ such that $\sigma(XY)=\sigma(X)\sigma(Y)$. The set of all such maps, denoted $\mathbf{Aut}(\mathbf{SE}_n(3))$ is a Lie group. By Lemma 2.1 in \cite{van2023constructiveINS}, $\sigma_Z$ is an automorphism of $\mathbf{SE}_n(3)$; i.e. $\sigma_Z \in \mathbf{Aut}(\mathbf{SE}_n(3))$.
\\ For any $A,B\in \mathbb{R}^{m \times n}$, the matrix commutator is given by
\begin{align*}
    [A,B] = AB-BA \ . 
\end{align*}
For all $A,B \in \mathbb{R}^{n\times m}$, the Euclidean inner product and the norm are defined by 
\begin{align*}
    \langle A,B\rangle = \text{tr}(A^\top B), && |A|^2 = \text{tr} (A^\top A),
\end{align*}
where tr: $\mathbb{R}^{m\times m}\rightarrow \mathbb{R}$. For the positive definite matrix $P\in \mathbb{R}^{m\times m}$ and $A\in \mathbb{R}^{n\times m}$, define the weighted norm
\begin{align*}
    |A|_P^2 = \langle A, AP\rangle = \text{tr}(APA^\top).
\end{align*}
$\mathbf{1}_n\in \mathbb{R}^n$ and $\mathbf{0}_n\in \mathbb{R}^n$ are column vectors with all 1's and 0's, respectively. 

\section{Problem Description}\label{sec:problem_description}
\subsection{LI-SLAM Dynamics and Measurements}
We consider mobile robot equipped with an Inertial Measurement Unit (IMU) and a 3D (e.g. RGBD or stereo) camera system, moving in an environment with static landmarks. 
The attitude, velocity, and position of the robot are denoted $R\in \mathbf{SO}(3)$, $v \in \mathbb{R}^3$, $x \in \mathbb{R}^3$, respectively, with respect to an arbitrary inertial frame $\{0\}$. 
The positions of the landmarks are denoted by $p_i$ in the same inertial frame, where $i= 1, \cdots,n$. 
The raw coordinates of the LI-SLAM problem are thus written as $(R,v,x,p_i) \in \mathbf{SO}(3)\times\mathbb{R}^3\times\mathbb{R}^3\times(\mathbb{R}^3)^n$. 
The state space of LI-SLAM problem, which we refer to as the \emph{total space}, is thus defined as
\begin{align}
    \mathcal{T}_n^{LI}(3) = \mathbf{SO}(3)\times(\mathbb{R}^3)^{n+2}.
\end{align}
The above formulation is an extension of the formulation presented in \cite{mahony2017geometric}, to also include velocity. 

The robot's onboard IMU provides measurements of its angular velocity $\Omega \in \mathbb{R}^3$ and proper acceleration $a \in \mathbb{R}^3$ in its body-fixed frame $\{B\}$.
The system dynamics are
\begin{align} \label{eq:system_dynamics}
    \dot{R} = R\Omega^\times, &&
    \dot{v} = Ra+g\mathbf{e}_3, &&
    \dot{x} = v, &&
    \dot{p}_i = 0,
\end{align}
for each $i = 1,..., n$, and where $g\mathbf{e}_3 \in \mathbb{R}^3$ is the gravity vector in the inertial frame (typically $g \approx 9.81 \text{m/s}^2$).

The 3D camera system provides measurements of the positions of the landmarks in the body frame, 
\begin{equation} \label{meas}
    y_i = h_i(R,v,x,p_i) = R^\top(p_i-x),
\end{equation}
for each $i = 1,..., n$.
For simplicity, we assume that all landmarks are measured at all times.
The notation $(R,v,x,p_i)\equiv (R,v,x,p_1, \cdots,p_n)$ is used for simplicity in the sequel.

\subsection{LI-SLAM invariance}

The LI-SLAM problem has a natural invariance associated with it.
Two trajectories of the LI-SLAM system are indistinguishable if they are related by a translation and rotation about the vertical $\mathbf{e}_3$ axis of the inertial reference frame \citep[Theorem 1]{martinelli2013observability}.
We demonstrate here how this invariance leads to a quotient manifold structure.

As in \citep{van2023eqvio}, the invariance of the problem is encoded by an action of the isotropy subgroup of $\mathbf{SE}(3)$ defined as: 
\begin{align}\label{eq:invariance_group_dfn}
    \mathbf{SE}_{\mathbf{e}_3}(3):= \{(R,x)\in \mathbf{SE}(3)|R\mathbf{e}_3 = \mathbf{e}_3\}.
\end{align}
This group acts on the total space by a group action
\begin{align}\label{eq:group_action_alpha}
    \alpha &: \mathbf{SE}_{\mathbb{e}_3}(3)\times \mathcal{T}_n^{LI}(3) \to \mathcal{T}_n^{LI}(3), \notag\\
    \alpha&(S,  (R,v,x,p_i)) := (R_S^\top R, R_S^\top v, R_S^\top(x-x_S),R_S^\top(p_i-x_S)).
\end{align}
This is a proper right group action of $\mathbf{SE}_{\mathbf{e}_3}(3)$ on $\mathcal{T}_n^{LI}(3)$. 
For a given $S\in \mathbf{SE}_{\mathbf{e}_3}(3)$, the action $\alpha(S, \cdot)$ represents a change of reference frame from \{0\} to \{1\}, where S is the pose of \{1\} with respect to \{0\}. 
The frame transformation by $S\in \mathbf{SE}_{\mathbf{e}_3}(3)$ leaves the gravity direction $\mathbf{e}_3$ unchanged. 
The dynamics \eqref{eq:system_dynamics} and measurements \eqref{meas} are invariant with respect to $\alpha$ \citep[Section IV.C]{van2023eqvio}. 

The group action $\alpha$ leads us to define a quotient manifold structure.
Given $(R,v,x,p_i)\in \mathcal{T}_n^{LI}(3)$, we define the equivalence class
\begin{align}
    [R,v,x,p_i] := \cset{\alpha(S,(R,v,x,p_i)) }{ S\in \mathbf{SE}_{\mathbf{e}_3}(3) }.
\end{align}
We refer to such an equivalence class $[R,v,x,p_i]$ as an \emph{LI-SLAM configuration}.
This leads to the set of all LI-SLAM configurations viewed as the quotient manifold of $\mathcal{T}_n^{LI}(3)$ under the action $\alpha$,
\begin{align}
   \mathcal{T}_n^{LI}(3)/\alpha 
   := \cset{[R,v,x,p_i]}{(R,v,x,p_i)\in \mathcal{T}_n^{LI}(3)},
\end{align}
with the associated quotient projection given by
\begin{align*}
    \zeta &: \mathcal{T}_n^{LI}(3) \to \mathcal{T}_n^{LI}(3)/\alpha, \\
    \zeta&(R,v,x,p_i) := [R,v,x,p_i].
\end{align*}
Then $\zeta$ is a smooth surjective submersion \citep[Theorem 21.10]{2012_lee_IntroductionSmoothManifolds}. 

Every LI-SLAM configuration $[R,v,x,p_i]$ is identified uniquely by the direction of gravity $R^\top \mathbf{e}_3$, the velocity $R^\top v$ and the relative landmark positions $R^\top(p_i - x)$ expressed in the body frame.
Formally, we define the LI-SLAM base space to be
\begin{align}\label{eq:LI-SLAM_base_space}
    \mathcal{M}_n^{LI}(3) := \mathbb{S}^2\times (\mathbb{R}^3)^{n+1}
\end{align}
along with a projection map
\begin{align}\label{eq:base_space_projection}
    \pi &: \mathcal{T}_n^{LI}(3)\rightarrow \mathcal{M}_n^{LI}(3), \notag \\
    \pi&(R,v,x,p_i) := (R^\top\mathbf{e}_3, R^\top v, R^\top(p_i-x)).
\end{align}
Then $\pi$ is a smooth surjective submersion from $\mathcal{T}_n^{LI}(3)$ to $\mathcal{M}_n^{LI}(3)$. 
Additionally, $\pi$ is invariant under $\alpha$, i.e. $\pi \circ (\alpha(S, \cdot)) = \pi( \cdot)$ for all $S\in \mathbf{SE}_{\mathbf{e}_3}(3)$, and thus induces a smooth map $\bar{\pi}:\mathcal{T}_n^{LI}(3)/\alpha \rightarrow \mathcal{M}_n^{LI}(3)$.
The manifolds we have discussed and the mappings between them are summarised in the following commutative diagram.
\[
\begin{tikzcd}
 \mathcal{T}_n^{LI}(3) \arrow{r}{\zeta} \arrow[swap]{dr}{\pi} & \mathcal{T}_n^{LI}(3)/\alpha \arrow{d}{\bar{\pi}} \\
 & \mathcal{M}_n^{LI}(3)
\end{tikzcd}
\]
The map $\bar{\pi}$ is a diffeomorphism, therefore allowing us to perform observer analysis in the base space $\mathcal{M}_n^{LI}(3)$ that contains all the observable states, and subsequently make inferences about the total space $\mathcal{T}_n^{LI}(3)$.

\subsection{Lie Group Interpretation}
\label{sec:lie-group-interpretation}

The states of the LI-SLAM system $(R,v,x,p_i) \in \mathcal{T}_n^{LI}(3)$ can be compactly represented in a matrix form using the Lie group $\mathbf{SE}_{n+2}(3)$ by writing
\begin{align} \notag
    X &= \begin{pmatrix}
        R & V\\
        0_{(n+2)\times 3}& I_{n+2} 
    \end{pmatrix} \in \mathbf{SE}_{n+2}(3), \\
    V &= \begin{pmatrix}
        v & x & p_1 & \cdots & p_{n}
    \end{pmatrix}\in \mathbb{R}^{3\times (n+2)}.
    \label{eq:lie-group_notation}
\end{align}
The translational sub-matrix $V$ encodes the velocity and position of the robot, and the positions of the landmarks.

Using the Lie group notation \eqref{eq:lie-group_notation}, the dynamics \eqref{eq:system_dynamics} can be written as 
\begin{align}\label{eq:lie-group_dynamics}
    \dot{X} = XU + GX + [N,X], 
\end{align}
where $U,G \in \mathfrak{se}_{n+2}(3), N\in \mathfrak{sim}_{n+2}(3)$, and
\begin{align*}
    U &= \begin{pmatrix}
        \Omega^\times & W_U\\
        0_{(n+2)\times3}& 0_{(n+2)\times(n+2)}
    \end{pmatrix}, & 
    W_U &= \begin{pmatrix}
        a & 0_{3\times(n+1)}
    \end{pmatrix}, \\
    G&= \begin{pmatrix}
        0_{3\times 3} & W_G\\
        0_{(n+2)\times3}& 0_{(n+2)\times(n+2)}
    \end{pmatrix}, &
    W_G &= \begin{pmatrix}
        g\mathbf{e}_3 &  0_{3\times(n+1)}
    \end{pmatrix} ,\\
    N &= \begin{pmatrix}
        0_{3 \times 3} & 0_{3 \times (n+2)}\\
        0_{(n+2) \times 3} & S_N
    \end{pmatrix} , &
    S_N &= \begin{pmatrix}
        0 & -1 & \mathbf{0}_n^\top\\
        0& 0 & \mathbf{0}_n^\top\\
        \mathbf{0}_n & \mathbf{0}_n & 0_{n\times n}
    \end{pmatrix} .
\end{align*}
We note the similarity to the representation of inertial navigation system dynamics presented in \cite{VANGOOR2025112328}.

The measurements can also be compactly written using the Lie group notation \eqref{eq:lie-group_notation}.
Specifically, one has that
\begin{align}\label{eq:lie-group_measurements}
    Y &= (y_1\ \cdots\ y_n) \in \mathbb{R}^{3\times n}, \notag \\
    \begin{pmatrix}
        Y \\ C
    \end{pmatrix}
    &= X^{-1} \begin{pmatrix}
        0_{3\times n} \\
        C
    \end{pmatrix}
    = \begin{pmatrix}
        - R^\top V C \\ C
    \end{pmatrix},
\end{align}
where
\begin{align*}
    C &= \begin{pmatrix}
        \mathbf{0}_n^\top \\
        \mathbf{1}_n^\top \\
        -I_n
    \end{pmatrix}
    \in \mathbb{R}^{(n+2)\times n}.
\end{align*}
In coordinates, the projection $\pi$ \eqref{eq:base_space_projection} from the total space $\mathcal{T}^{LI}_n(3)$ to the base space $\mathcal{M}^{LI}_n(3)$ can be written as
\begin{align*}
    \pi(X) &= (R^\top \mathbf{e}_3, -R^\top V \Pi), \\
    \Pi &:= \begin{pmatrix}
        -1 & \mathbf{0}_n^\top \\
        0 & \mathbf{1}_n^\top \\
        \mathbf{0}_{n} & -I_n
    \end{pmatrix} \in \mathbb{R}^{(n+2)\times(n+1)}.
\end{align*}
Here the first column of $\Pi$ extracts the body-fixed velocity $R^\top v$ while the remaining columns extract the $n$ body-fixed landmark positions $R^\top (p_i - x)$.
Using the matrix $\Pi$, the measurement matrix $Y$ can also be written as
\begin{align} 
    Y &= (-R^\top V \Pi) C', &
    C' &= \begin{pmatrix}
        \mathbf{0}_n^\top \\ I_n
    \end{pmatrix} \in \mathbb{R}^{(n+1)\times n}.
    \label{eq:Y_matrix_dfn}
\end{align}

In summary, the dynamics, measurements, and invariance of LI-SLAM all admit compact and natural representations by using the Lie group structure of $\mathbf{SE}_{n+2}(3)$.
We exploit these representations in the observer design in Section \ref{sec:observer_design}.
In particular, we identify the total space with the Lie group $\mathcal{T}_{n+2}^{LI}(3) \simeq \SE_{n+2}(3)$ throughout the remainder of the paper.

\section{Observer Design}\label{sec:observer_design}

\subsection{Synchronous Observer Architecture}
 
We apply the observer architecture proposed in \citep{van2021autonomous,van2025synchronous} by exploiting $\SIM_{n+2}(3)$ to represent the automorphisms of $\SE_{n+2}(3)$.
The observer state is defined as $\hat{X} = (\hat{R},\ \hat{V}) \in \mathbf{SE}_{n+2}(3)$, where $\hat{V} = (\hat{v}\ \hat{x}\ \hat{p}_i)$, and the auxiliary state is defined as $Z = (R_Z,V_Z, A_Z) \in \mathbf{SIM}_{n+2}(3)$, where $V_Z = (v_Z\ x_Z\ p_{Zi})$.
Following the general architecture in \cite{van2021autonomous}, their dynamics are defined by
\begin{align} \label{eq:observer_arch}
    \dot{\hat{X}} &= \hat{X}U + G\hat{X} + [N,\hat{X}] + Z\Delta Z^{-1} \hat{X},\notag \\
     \dot{Z} &= (G+N)Z - Z\Gamma ,
\end{align}
where $\Delta \in \mathfrak{se}_{n+2}(3)$ and $\Gamma \in \mathfrak{sim}_{n+2}(3)$ are Lie-algebra valued correction terms to be designed later. 

Define the observer error in the total space 
\begin{align} \label{eq:total_space_error}
    \bar{E} := \sigma_Z^{-1}(X\hat{X}^{-1}) = Z^{-1}X\hat{X}^{-1}Z.
\end{align}
Then the dynamics of error $\bar{E}$ are given by
\begin{align}\label{eq:total_error_dyn}
    \dot{\bar{E}} = \Gamma\bar{E} - \bar{E}\Gamma- \bar{E}\Delta .
\end{align}
Thus, the observer and the system are $\bar{E}$-synchronous. 
That is, the error dynamics depends only on the chosen correction terms $\Delta$ and $\Gamma$, and $\dot{\bar{E}}=0$ if the correction terms are set to zero.

The observer structure in \eqref{eq:observer_arch} involves two dynamical systems - one in the observer variable $\hat{X} \in \mathbf{SE}_{n+2}(3) $, and the other in the auxiliary variable $Z
\in \mathbf{SIM}_{n+2}(3)$ that affects the observer dynamics.
The latter involves $9 + 3n + (n+2)^2$ independent variables
and an identical number of initial conditions, including 3 for the rotation $R_Z \in \mathbf{SO}(3)$, $3(n+2)$ for the translation $V_Z \in \mathbb{R}^{3\times (n+2)}$, and $(n+2)^2$ for the scaling matrix $A_Z \in \mathbf{GL}(n+2)$
These are essentially dynamic design parameters that are chosen to meet the stability, the performance, and in addition, the synchrony requirements of the observer.

Although the observer architecture allows for any $Z\in \mathbf{SIM}_{n+2}(3)$, in our current objective we choose correction terms $\Gamma$ that render $Z$ constant, thereby simplifying the resulting design.
Nonetheless, the introduction of $Z$ is vital to the synchronous observer design methodology and the analysis of the observer's error dynamics.
Let $R_Z(0)=I_3$ and $A_Z(0) = I_{n+2}$, and choose $\Omega_\Gamma = 0$ and $S_\Gamma=S_N$. Then $\dot{R}_Z = 0$ and $\dot{A}_Z = 0$, and therefore $R_Z\equiv I_3$ and $A_Z\equiv I_{n+2}$ for all time.
It follows that $Z=(I_3,V_Z,I_{n+2})$, and $R_Z$ and $A_Z$ will not be considered in the sequel.
The translation components $V_Z$ will also be chosen constant, although the value of this constant will depend on the chosen gains for the observer (see Theorem \ref{th:observer_thm}).

Considering $Z = (I_3, V_Z, A_Z)$, if $\bar{E} = (R_{\bar{E}}, V_{\bar{E}}) \in \mathbf{SE}_{n+2}(3)$, then the rotational error $(R_{\bar{E}})$ and the translational error $(V_{\bar{E}})$ are computed as
\begin{align}\label{eq:total_error_components}
    R_{\bar{E}} = R\hat{R}^\top, && V_{\bar{E}} = (V- R_{\bar{E}}\hat{V}) - (I_3 - R_{\bar{E}})V_Z.
\end{align}
The rotation error and the first term in the translation error are familiar as the classic right-invariant Lie group errors.
In our synchronous observer design, we additionally have the term $(I_3 - R_{\bar{E}})V_Z$ as a result of the conjugation by $Z$ in \eqref{eq:total_space_error}, which is necessary for compensating the right-invariant term $GX$ and the group-linear term $[N,X]$ in the dynamics \eqref{eq:lie-group_dynamics}.
The projection of $\bar{E}$ from the total space to the base space is given by $\bar{e} = \pi(\bar{E}) = (\eta_e, V_e^o) \in \mathcal{M}_n^{LI}(3)$, where
\begin{align} \label{eq:base_space_error}
    \eta_e = R_{\bar{E}}^\top\mathbf{e}_3, \; V_e^o = -(R_{\bar{E}}^\top V - \hat{V})\Pi + (R_{\bar{E}}^\top-I_3)V_Z\Pi.
\end{align}
Due to the unobservability of the system posed on the total space, the observer design will guarantee only that the projected error $\bar{e}$ converges to the origin $(\mathbf{e}_3, 0_{3\times (n+1)})$, and that the total space error $\bar{E}$ converges to a constant that is not necessarily the identity.

\subsection{Observer Design}

The following theorem defines the correction terms used to ensure convergence (up to reference frame invariance) of the state estimate $\hat{X}$ to the state $X$.
The auxiliary state $Z$ introduced in the previous section is an important part of the architecture and is required to attain synchronous error dynamics.
As we have shown, however, by making specific choices for the correction terms $\Omega_\Gamma$ and $S_\Gamma$, the components of $R_Z$ and $A_Z$ $Z$ can be made constant.
In the theorem below, we additionally choose a correction term $W_\Gamma$ and an initial condition $V_Z(0)$ that also makes $V_Z$ constant for all time.
This means that the only dynamic state in the observer is the state estimate $\hat{X}$.
This makes the observer \emph{minimal} in the sense that the state space of its dynamics is exactly the state space of the original system, and there are no auxiliary or virtual states with dynamics that need to be tracked.

\begin{thm}\label{th:observer_thm}    
    Let $X \in \SE_{n+2}(3)$ denote the LI-SLAM state with dynamics \eqref{eq:system_dynamics} and measurements \eqref{meas}, and let $\hat{X}\in \mathbf{SE}_{n+2}(3)$ denote the observer state with dynamics \eqref{eq:observer_arch}.
    The estimated measurements are written as $\hat{y}_i = h_i(\hat{X}) = \hat{R}^\top(\hat{p}_i-\hat{x})$ and $\hat{Y}=(\hat{y}_1 \ \cdots\ \hat{y}_n)$.
    Choose gains $k_R > 0$, $k_p + nk_x>0$ and $k_p,k_v > 0$ and 
    Initialize the auxiliary state $Z \in \SIM_{n+2}(3)$ by
    \begin{align}\label{eq:constant_Z_dfn}
        Z &= (I_3,V_Z,I_{n+2}), \notag\\
        V_Z &= \begin{pmatrix}
            \frac{(k_p+nk_x)g}{nk_v} \mathbf{e}_3 &  \frac{g}{nk_v} \mathbf{e}_3& 0_{3\times n}
    \end{pmatrix}.
    \end{align}
    Define the correction terms $\Delta = (\Omega_\Delta, W_\Delta)\in \mathfrak{se}_{n+2}(3)$ and $W_\Gamma \in \mathbb{R}^{3\times(n+2)}$ by
    \begin{align}\label{eq:correction_terms}
            \Omega_\Delta &= k_R\ \mathbf{e}_3^\times \hat{R}(Y - \hat{Y}) \mathbf{1}_n, \\
            W_\Delta &= \hat{R}(Y - \hat{Y})K,  \\
            W_\Gamma &= -V_Z(CK+K_Z), \\
            K &= \begin{pmatrix}
                -k_v\mathbf{1}_n & -k_x\mathbf{1}_n & k_pI_n 
            \end{pmatrix}
            \in \mathbb{R}^{n \times (n+2)} \notag \\
            K_Z &= \begin{pmatrix}
    0 & 0 &\mathbf{0}_{n}^\top \\
    0 & -k_p & -k_p\mathbf{1}_n^\top\\
    \mathbf{0}_{n} & \mathbf{0}_{n} & 0_{n\times n}
\end{pmatrix} \in \mathbb{R}^{(n+2)\times(n+2)}. \notag
    \end{align}
    Let the total space error $\bar{E}$ and base space error $\bar{e} = (\eta_e, V_e^o)$ be defined as in (\ref{eq:total_space_error},\ref{eq:base_space_error}).
    Then
    \begin{enumerate}
        \item The auxiliary state $Z$ remains constant.
        \item The origin of the translational error $V_e^o$ is globally exponentially stable.
        \item The reduced attitude error $\eta_e$ is almost-globally asymptotically and locally exponentially stable to $\mathbf{e}_3$, with its only unstable equilibrium at $-\mathbf{e}_3$.
        \item If the base space error $\pi(\bar{E}) = \bar{e}$ converges to the origin $(\mathbf{e}_3, 0_{3\times n+1})$, then the total space error $\bar{E}$ converges to a constant, and the base space projection of the estimated state converges to that of the true state, that is, $\pi(\hat{X}) \to \pi(X)$.
    \end{enumerate}
\end{thm}

%\pieter{Check the rewrite}
The observer design in Theorem \ref{th:observer_thm} is given in terms of the matrix notation introduced in Section \ref{sec:lie-group-interpretation}, but can also be written in terms of its state components $\hat{X} = (\hat{R}, \hat{v}, \hat{x}, \hat{p}_i)$.
A straightforward expansion of the correction terms \eqref{eq:correction_terms} and observer dynamics \eqref{eq:observer_arch} yields simplified equations,
\begin{align*}
    \dot{\hat{R}} &= \hat{R}\Omega^\times + \Omega_\Delta^\times\hat{R}, \qquad
    \Omega_\Delta = k_R \mathbf{e}_3 \times \sum_{i=1}^n \hat{R}(y_i-\hat{y}_i),
    \\
    \dot{\hat{v}} &= \hat{R}a + g\mathbf{e}_3 -k_v\sum_{i=1}^n\hat{R}(y_i-\hat{y}_i) + \Omega_\Delta^\times(\hat{v} - v_Z), \\
    \dot{\hat{x}} &= \hat{v} -k_x\sum_{i=1}^n\hat{R}(y_i-\hat{y}_i) + \Omega_\Delta^\times(\hat{x} - x_Z), \\
    \dot{\hat{p}}_i &= k_p\hat{R}(y_i-\hat{y}_i) + \Omega_\Delta^\times\hat{p}_i,
\end{align*}
where $v_Z = \frac{(k_p+nk_x)g}{nk_v}\mathbf{e}_3$ and $x_Z = \frac{g}{nk_v}\mathbf{e}_3$ are constants.

\begin{proof}
\underline{Proof of item 1}):
We begin by showing that the constant $Z \in \SIM_{n+2}(3)$ defined in \eqref{eq:constant_Z_dfn} satisfies the auxiliary state dynamics \eqref{eq:observer_arch} for the particular choice of $W_\Gamma$ in Theorem \ref{th:observer_thm}.
The dynamics of $A_Z$ and $R_Z$ were already chosen such that $A_Z = I_{n+2}$ and $R_Z = I_3$ for all time.
The dynamics of $V_Z$ are
\begin{align*}
    \dot{V}_Z &= W_G-W_\Gamma-V_ZS_N \\
    &= W_G+V_Z(CK+K_Z)-V_ZS_N\\
    &= W_G+V_Z(CK+K_Z - S_N).
\end{align*}
The matrix $(CK+K_Z-S_N)$ has full row rank and is thus invertible, meaning that $V_Z=-W_G(CK+K_Z-S_N)^{-1}$ is a constant solution to the dynamics.
This is exactly the initial value of $V_Z$ stated in \eqref{eq:constant_Z_dfn}.

\underline{Proof of item 2}):
To show the stability of $V_e^o = -R_{\bar{E}}^\top V_{\bar{E}} \Pi$, we begin by examining the dynamics of $\bar{E}$.
Expanding \eqref{eq:total_error_dyn}, the dynamics of $V_{\bar{E}}$ and $R_{\bar{E}}$ are
\begin{subequations}
    \begin{align}
    \dot{R}_{\bar{E}} &= -R_{\bar{E}}\Omega_\Delta^\times,\\
    \dot{V}_{\bar{E}} &= -V_{\bar{E}}S_N + (I-R_{\bar{E}})W_{\Gamma} - R_{\bar{E}}W_\Delta .
\end{align}
\end{subequations}
Recalling \eqref{eq:lie-group_measurements} and \eqref{eq:total_error_components}, the correction term $W_\Delta$ can be expanded as
\begin{align*}
    W_\Delta &= \hat{R} (Y - \hat{Y}) K \\
    &= \hat{R}(-R^\top V C + \hat{R}^\top \hat{V} C) K \\
    &= - R_{\bar{E}}^\top (V_{\bar{E}} + (I_3 - R_{\bar{E}})V_Z) C K \\
    &= - R_{\bar{E}}^\top V_{\bar{E}} C K - (R_{\bar{E}}^\top - I_3)V_Z C K.
\end{align*}
Thus, the dynamics of $V_e^o$ are
\begin{subequations}
\begin{align}
    \dot{V}_e^o 
    &= -\Omega_\Delta^\times R_{\bar{E}}^\top V_{\bar{E}}\Pi 
    + R_{\bar{E}}^\top V_{\bar{E}}S_N\Pi 
    - (R_{\bar{E}}^\top -I)W_{\Gamma}\Pi 
    + W_\Delta \Pi \notag\\
    % ------------------------
    &= \Omega_\Delta^\times V_e^o 
    + R_{\bar{E}}^\top V_{\bar{E}}S_N\Pi 
    + (R_{\bar{E}}^\top-I) V_Z (CK+K_Z)\Pi
    \notag\\&\quad
    + (- R_{\bar{E}}^\top V_{\bar{E}} C K - (R_{\bar{E}}^\top - I_3)V_Z C K) \Pi \notag\\
    % ------------------------
    &= \Omega_\Delta^\times V_e^o 
    + R_{\bar{E}}^\top V_{\bar{E}}S_N\Pi 
    - R_{\bar{E}}^\top V_{\bar{E}} C K \Pi 
    \label{eq:Veo_1} \\
    % ------------------------
    &= \Omega_\Delta^\times V_e^o 
    + R_{\bar{E}}^\top V_{\bar{E}} \Pi S_N' 
    - R_{\bar{E}}^\top V_{\bar{E}} \Pi C' K \Pi 
    \label{eq:Veo_2} \\
    % ------------------------
    &= \Omega_\Delta^\times V_e^o 
    + V_e^o (C' K\Pi - S_N'),\notag
    % ------------------------
\end{align}
\end{subequations}
where \eqref{eq:Veo_1} follows from $K_Z \Pi = 0$, and \eqref{eq:Veo_2} follows from $\Pi C' = C$ and $\Pi S_N' = S_N \Pi$, where 
\[
S_N' = \begin{pmatrix}
    0 & \mathbf{1}_n^\top \\
    \mathbf{0}_n & 0_{n\times n}
\end{pmatrix} \in \mathbb{R}^{(n+1)\times (n+1)}.
\] 
Let $A = C'K\Pi - S_N' \in \mathbb{R}^{(n+1) \times (n+1)}$, then its characteristic polynomial is
\begin{align*}
    \det(sI_{n+1}-A) = (s+k_p)^{n-1}(s^2 +(k_p+nk_x)s+nk_v),
\end{align*}
with the solutions
%\footnote{The solution $s = -k_p$ has multiplicity $n-1$.}
\begin{align*}
    s = \frac{-(k_p+nk_x) \pm \sqrt{(k_p+nk_x)^2-4k_v}}{2}, \; -k_p ,
\end{align*}
which have strictly negative real parts for the chosen gains $k_p,k_v>0$ and $k_p+nk_x>0$.
Hence $A$ is Hurwitz, and thus there exists a unique positive definite $P\in \mathbb{R}^{(n+1)\times(n+1)}$ satisfying the Lyapunov equation $AP+PA^\top = - I_{n+1}$.

Consider the candidate Lyapunov function for $V_e^o$,
\begin{align}\label{eq:v_cost_fn}
    \mathcal{L}_V = |V_e^o|_P^2 \ .
\end{align}
The derivative of $\mathcal{L}_V$ is given by 
\begin{align} \label{eq:L_V_dyn}
        \dot{\mathcal{L}}_V &= \text{tr}(\dot{V}_e^oP V_e^{o\top}) +  \text{tr}(V_e^oP \dot{V}_e^{o\top}) \notag\\
        &= \text{tr}((V_e^oA + \Omega_\Delta^\times V_e^{o})PV_e^{o\top}) + \text{tr}(V_e^oP(A^\top V_e^{o\top} - V_e^{o\top}\Omega_\Delta^\times)) \notag \\
        &= \text{tr}(V_e^o(AP+AP^\top)V_e^{o\top}) \notag \\
        % &= -\text{tr}(V_e^oQV_e^{o\top}),\\
        &= -|V_e^o|^2 .
\end{align}
Therefore, $V_e^o$ is indeed globally exponentially stable to zero.
    
\underline{Proof of item 3}): 
To study the stability of the reduced attitude $\eta_e$, consider the candidate Lyapunov function
\begin{align}\label{eq:lyapunov_function}
    \mathcal{L}(\eta_e,V_e^o):= \frac{1}{2}|\eta_e - \mathbf{e}_3|^2 + q|V_e^o|_P^2,
\end{align}
where $q = 2 n k_vk_R/g$.
The attitude correction term $\Omega_\Delta$ can be written as
\begin{align*}
    \Omega_\Delta 
    &= k_R\ \mathbf{e}_3^\times \hat{R}(Y - \hat{Y}) \mathbf{1}_n \\
    &= k_R\ \mathbf{e}_3^\times (- R_{\bar{E}}^\top V_{\bar{E}} C - (R_{\bar{E}}^\top - I_3)V_Z C) \mathbf{1}_n \\
    &= k_R\ \mathbf{e}_3^\times (V_e^o C' - (R_{\bar{E}}^\top - I_3) \frac{g}{nk_v}\mathbf{e}_3 \mathbf{1}_n^\top) \mathbf{1}_n \\
    &= k_R\ \mathbf{e}_3^\times (V_e^o C' \mathbf{1}_n  - (R_{\bar{E}}^\top - I_3) \frac{g}{k_v}\mathbf{e}_3) \\
    &= -\frac{k_Rg}{k_v} (\mathbf{e}_3^\times \eta_e)  + k_R\ \mathbf{e}_3^\times V_e^o C' \mathbf{1}_n.
\end{align*}
Thus the derivative of $\mathcal{L}$ is given by 
\begin{align*}
    \dot{\mathcal{L}} 
    &= \langle \eta_e- \mathbf{e}_3, \Omega_\Delta^\times \eta_e \rangle 
    -q |V_e^o|^2 \\
    % ---------------
    &= \langle \mathbf{e}_3^\times \eta_e, \Omega_\Delta\rangle 
    -q |V_e^o|^2 \\
    % ---------------
    &= \left\langle \mathbf{e}_3^\times \eta_e, -\frac{k_Rg}{k_v} (\mathbf{e}_3^\times \eta_e)  + k_R\ \mathbf{e}_3^\times V_e^o C' \mathbf{1}_n \right\rangle
     -q |V_e^o|^2 \\
    % ---------------
    &= -\frac{k_Rg}{k_v}|\mathbf{e}_3^\times\eta_e|^2 
    + k_R \langle\mathbf{e}_3^\times \eta_e, \mathbf{e}_3^\times V_e^o C' \mathbf{1}_n\rangle
    -q |V_e^o|^2 \\
    % ---------------
    &\le -\frac{k_Rg}{k_v}|\mathbf{e}_3^\times\eta_e|^2 
    + k_R \vert \mathbf{e}_3^\times \eta_e \vert \vert \mathbf{e}_3^\times V_e^o C' \mathbf{1}_n \vert
    -q |V_e^o|^2\\
    % ---------------
    &\le -\frac{k_Rg}{k_v}|\mathbf{e}_3^\times\eta_e|^2 
    + k_R \vert \mathbf{e}_3^\times \eta_e \vert \vert \mathbf{e}_3^\times \vert \vert V_e^o \vert \vert C' \mathbf{1}_n \vert
    -q |V_e^o|^2\\
    % ---------------
    &\le -\frac{k_Rg}{k_v}|\mathbf{e}_3^\times\eta_e|^2 
    + 2 k_R \sqrt{n} \vert \mathbf{e}_3^\times \eta_e \vert \vert V_e^o \vert
    - \frac{2 nk_vk_R}{g} |V_e^o|^2\\
    % ---------------
    &= -\frac{k_Rg}{k_v}(|\mathbf{e}_3^\times\eta_e| - \frac{\sqrt{n}k_v}{g}|V_e^o|)^2 - \frac{nk_vk_R}{g} |V_e^o|^2.
\end{align*}
The derivative of $\mathcal{L}$ is negative semi-definite with equality to zero only when $e_3^\times \eta_e = 0$ and $V_e^o = 0$. 
We know $\dot{\mathcal{L}}$ is uniformly continuous, as it is the composition of sums and products of uniformly continuous functions. 
By Barbalat's lemma (\cite{slotine1991nonlinear}, Lemma 4.2/4.3), $\mathcal{L} \rightarrow \mathcal{L}_{lim} \ge 0$ and $\dot{\mathcal{L}} \rightarrow 0$, where $\mathcal{L}_{lim} \le \mathcal{L}(\eta_e(0), V_e^o(0))$ is a constant. 
Since $V_e^o \rightarrow 0$ globally exponentially, $\mathcal{L} \rightarrow \frac{1}{2} |\eta_e - \mathbf{e}_3|^2 \rightarrow \mathcal{L}_{lim}$ and $\dot{\mathcal{L}} \rightarrow -\frac{k_Rg}{k_v}|\eta_e^\times\mathbf{e}_3|\rightarrow0$. 
Hence, $\eta_e^\times\mathbf{e}_3\rightarrow 0$ which implies that $\eta_e \rightarrow\mathbf{e}_3$ or $\eta_e \rightarrow -\mathbf{e}_3$. 
The equilibrium $(-\mathbf{e}_3,0)$ is unstable since $\mathcal{L}(-\mathbf{e}_3, 0) = 2$ is the global maximum of $\mathcal{L}(\eta_e,0)$.
The equilibrium $(\mathbf{e}_3, 0)$ is the unique global minimiser of $\mathcal{L}$, and therefore $\eta_e$ is almost-globally asymptotically stable.

Linearising the dynamics of $\eta_e$ about $\eta_e \approx \mathbf{e}_3+\varepsilon_\eta$, where $\varepsilon_\eta = (\varepsilon_1\ \varepsilon_2\ 0)^\top\in \mathbb{R}^3$ and $V_e^o \approx 0$, one has%
\footnote{The third component of $\varepsilon$ is zero since it must lie in the tangent space of the sphere at $\mathbf{e}_3$.}
\begin{align*}
    \dot{\varepsilon}_\eta &\approx -\frac{k_Rg}{k_v}\ (\mathbf{e}_3^\times\varepsilon_\eta)^\times\mathbf{e}_3, \\
    &= \frac{k_Rg}{k_v}\ \mathbf{e}_3^\times\mathbf{e}_3^\times \varepsilon_\eta, \\
    &= \frac{k_Rg}{k_v}\ (\mathbf{e}_3\mathbf{e}_3^\top-I_3)\varepsilon_\eta.
\end{align*}
In other words, the linearisation satisfies $\dot{\varepsilon}_1 \approx -\frac{k_Rg}{k_v}\ \varepsilon_1$ and $\dot{\varepsilon}_2 \approx -\frac{k_Rg}{k_v}\ \varepsilon_2$.
Therefore $\eta_e$ is locally exponentially stable.
    
\underline{Proof of item 4}): 
If $\bar{e} = (\eta_e, V_e^o) \to (\mathbf{e}_3, 0_{3\times n+1})$, then
\begin{align*}
    \hat{R}(Y-\hat{Y}) 
    &= - R_{\bar{E}}^\top V_{\bar{E}} C K - (R_{\bar{E}}^\top - I_3)V_Z \\
    &= - R_{\bar{E}}^\top V^o_{e} C' K + (R_{\bar{E}}^\top - I_3) W_G(CK+K_Z-S_N)^{-1} \\
    &\to - R_{\bar{E}}^\top 0_{3\times (n+1)} C' K + 0_{3\times (n+2)} (CK+K_Z-S_N)^{-1} \\
    &= 0_{3\times n}.
\end{align*}
It follows that the correction terms $\Omega_\Delta, W_\Delta \to 0$ exponentially, and $R_{\bar{E}}W_{\Gamma} \to W_\Gamma$.
Thus, the attitude error dynamics $\dot{R}_{\bar{E}} \to 0$ exponentially and the translation error dynamics become
\begin{align*}
    \dot{V}_{\bar{E}} 
    &= -V_{\bar{E}}S_N + (I-R_{\bar{E}})W_{\Gamma} - R_{\bar{E}}W_\Delta \\
    &\to -V_{\bar{E}}S_N \\
    &= -\begin{pmatrix}
        v_{\bar{E}} & x_{\bar{E}} & p_{\bar{E}, 1} & \cdots p_{\bar{E}, n}
    \end{pmatrix}S_N \\
    &= \begin{pmatrix}
        \mathbf{0}_3 & v_{\bar{E}} & 0_{3\times n}
    \end{pmatrix} \\
    &\to 0_{3\times n+2},
\end{align*}
also exponentially fast, where the last line is due to the fact that $R_{\bar{E}}^\top v_{\bar{E}} \to 0$ as a consequence of $V_e^o \to 0$.
Thus, $\dot{\bar{E}} \to 0$ exponentially fast and therefore $\bar{E}$ converges to a constant belonging to the invariance group $\mathbf{SE}_{\mathbf{e}_3}(3)$, as $R_{\bar{E}}^\top\mathbf{e}_3 \to \mathbf{e}_3$ and $x_{\bar{E}}-p_{\bar{E},i} \to 0_{3\times 1}$.

Finally, to see that $\pi(\hat{X}) =: (\hat{R}^\top \mathbf{e}_3, - \hat{R}^\top \hat{V} \Pi)$ converges to $\pi(X) =: (R^\top \mathbf{e}_3, -R^\top V \Pi)$, we have that
\begin{align}
    \hat{R}^\top \mathbf{e}_3 
    &= R^\top R \hat{R}^\top \mathbf{e}_3 
    = R^\top R_{\bar{E}} \mathbf{e}_3 
    \to R^\top \mathbf{e}_3,\notag \\
    - \hat{R}^\top \hat{V} \Pi
    &= - R^\top R_{\bar{E}} \hat{V} \Pi \notag\\
    &= - R^\top R_{\bar{E}} \left(
        R_{\bar{E}}^\top V -(R_{\bar{E}}^\top V - \hat{V})
        % + (R_{\bar{E}}^\top - I_3) V_Z
    \right) \Pi \notag\\
    &\to - R^\top R_{\bar{E}} \left(
        R_{\bar{E}}^\top V \Pi + V_e^o
    \right) \label{eq:convergence}\\
    &\to - R^\top R_{\bar{E}} R_{\bar{E}}^\top V \Pi \notag\\
    &= - R^\top V \Pi, \notag
\end{align}
where \eqref{eq:convergence} follows from \eqref{eq:base_space_error} and $(R_{\bar{E}}^\top-I_3)V_Z\Pi \to 0_{3\times 1}$ as the constant $V_Z$ is of the form $\mathbf{e}_3L,\ L\in \mathbb{R}^{1\times(n+2)}$ and $R_{\bar{E}}^\top\mathbf{e}_3 \to \mathbf{e}_3$.
Therefore, indeed, the base space projection of the estimated state converges to that of the true state.
\end{proof}

\section{Simulations}\label{sec:simulations}
\begin{figure}[h!]
    \centering
    \includegraphics[trim=0cm 0cm 0cm 0cm, clip, width=0.6\textwidth]{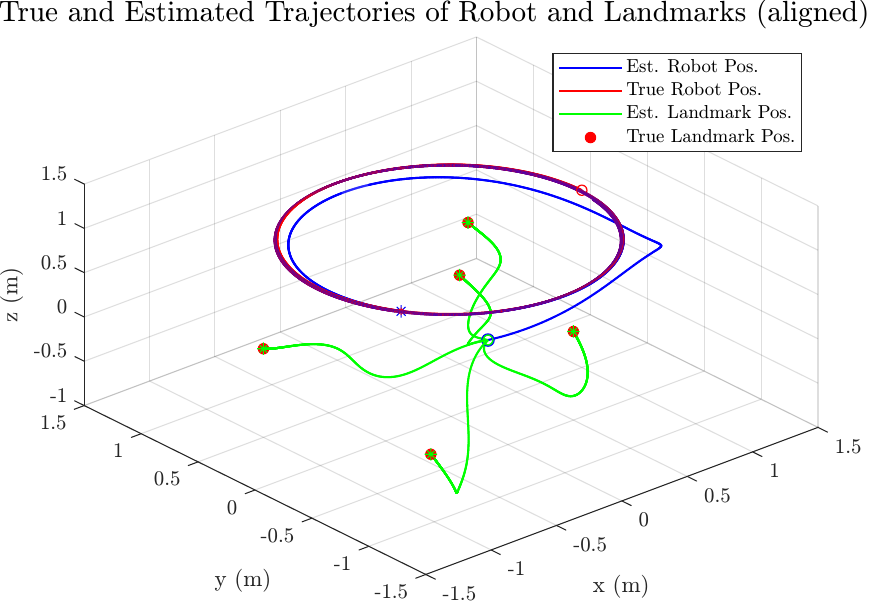}
    \caption{The trajectories of the true and estimated robot and landmark positions over time. 
    All the initial and final positions are marked with $\circ$ and $*$ respectively.
    The trajectories are aligned as described in Section \ref{sec:simulations}
    }
    \label{fig:traj}
\end{figure}

To verify the proposed observer, we simulated a robot flying uniformly in a circular trajectory of radius 1 m at a height of 1 m, viewing five static landmarks on the ground.
The true robot states were initialised as
\begin{align*}
    R(0) = I_3, && v(0)= \mathbf{e}_2\ \text{m/s}, && x(0)=(1\ 0\ 1)^\top\text{m}.
\end{align*}
The five landmarks were defined as $p_1=(0.5\ 0.5\ 0)^\top,\ p_2=(0.5\ -0.5\ 0)^\top,\ p_3 = (-1\ 0.5\ 0)^\top,\ p_4=(1\ 1\ 0)^\top$ and $p_5=(-1.2\ -1.2\ 0)^\top$. 
The input signals were chosen as
\begin{align*}
    \Omega(t)= (0\ 0\ 1)^\top,&& 
    a(t) = (-1\ 0\ -g)^\top.
\end{align*}
The observer states were initialised as 
\begin{align*}
    &\hat{R}(0) = \exp(0.25\pi \mathbf{a}^\times),
    &&\hat{v}(0) = (0\ 0\ 0)^\top,\\
    &\hat{x}(0) = (0\ 0\ 0)^\top,
    &&\hat{p}_i(0) = (0\ 0\ 0)^\top,
\end{align*}
where $\mathbf{a} = (1\ 1\ 1)^\top$ and $i= 1,\ldots,5$. 
The gains were chosen to be $k_v = 2.0$, $k_x=1.0$, $k_p = 4.0$ and $k_R = 2.0$. Both the system and observer equations were simulated for 10~s using Euler integration at 500 Hz.
%\pieter{Review the rewrite}
Figure~\ref{fig:traj} shows the true and the \emph{aligned} estimated trajectories of the robot and the landmark positions.
Due to the invariance $\alpha$, the estimated state converges to the true state up to a constant rotation about yaw and a constant translation of the reference frame.
As $\pi(\hat{X}) \to \pi(X)$, the rotation and translation errors $R_{\bar{E}}, x_{\bar{E}}$ at the final time approximate this offset in $\mathbf{SE}_{\mathbf{e}_3}(3)$.
Thus, to show clearly the convergence of the observer to the true state, we realigned the estimated trajectory by transforming it with $\alpha(S^{-1}, \hat{X})$, where $S = (\exp{(\theta\mathbf{e}_3^\times)}, x_{\bar{E}})$, and $\theta = \operatorname{atan2}(R_{\bar{E},21}, R_{\bar{E},11}) \in [0, 2\pi)$ is the yaw angle of the rotation error $R_{\bar{E}}$.

\begin{figure}[h!]
    \centering
    \includegraphics[trim=0cm 0cm 0cm 0cm, clip, width=0.48\textwidth]{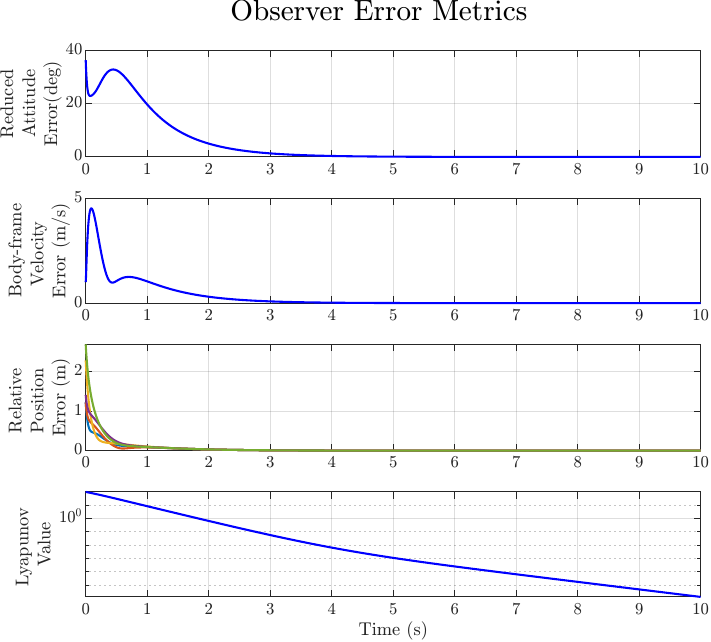}
    \caption{The reduced attitude error, body frame velocity error and the errors in relative landmark positions in body frame converge to zero. 
    The value of the Lyapunov function decreases steadily over time.
    }
    \label{fig:error}
\end{figure}

\begin{figure}[h!]
    \centering
    \includegraphics[trim=0cm 0cm 0cm 0cm, clip, width=0.6\textwidth]{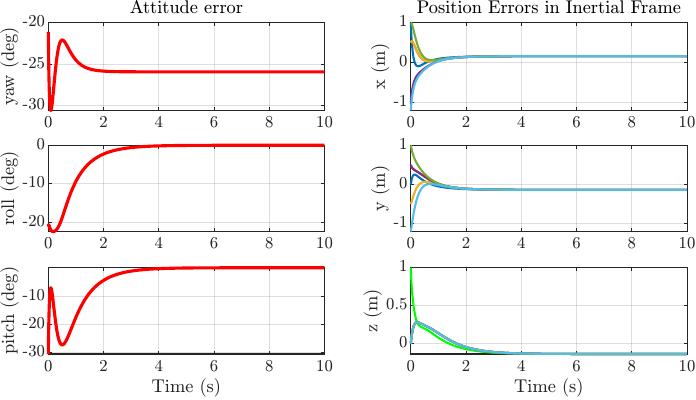}
    \caption{The attitude errors in roll and pitch converge to zero while the attitude error in yaw converges to a constant.
    The errors in robot and landmark positions also converge to a constant.
    }
    \label{fig:err_inertial}
\end{figure}

Figure~\ref{fig:error} shows the evolution of the reduced attitude error $(\arccos{(\mathbf{e}_3^\top\eta_e)})$, body frame velocity error $(|R^\top v - \hat{R}^\top\hat{v}|)$, relative landmark position errors $(|R^\top(p_i-x)-\hat{R}^\top(\hat{p}_i-\hat{x})|)$, and the value of the Lyapunov function \eqref{eq:lyapunov_function} over time. 
The value of the Lyapunov function decreases monotonically all the time, as expected from the proof of Theorem~\ref{th:observer_thm}. 
Figure~\ref{fig:err_inertial} shows that in the total space $\mathcal{T}_n^{LI}(3)$, attitude errors in roll and pitch go to zero and the error in yaw converges to a constant. 
It also shows that the errors in the robot position $(x-R_{\bar{E}}\hat{x})$ and the landmark positions $(p-R_{\bar{E}}\hat{p})$ in the inertial frame go to the same constant.
\section{Conclusion}\label{sec:conclusion}
This paper presents an observer design for landmark-inertial SLAM based on recent developments in the synchronous observe design methodology for group-affine systems \citep{van2021autonomous, VANGOOR2025112328}. 
The problem is analysed in terms of the observable base space using a Lie group action to represent the invariance of the LI-SLAM problem, motivated by the SLAM invariance introduced in \citep{mahony2017geometric}. 
The resulting observer is the first solution to LI-SLAM with almost-global asymptotic and local exponential stability that is minimal, in the sense that the state space of the observer is exactly the same as that of the state to be estimated, which was achieved by cancelling the dynamics of the auxiliary state to render it constant.
Finally, simulations demonstrate its almost-global asymptotic stability by showing that the observer states converge even from poor initial estimates.
This work contributes to the ongoing development of nonlinear observer-based SLAM frameworks, which are of particular interest due to their provable almost-global convergence properties and favorable computational efficiency compared to the state-of-art optimization- and EKF-based SLAM systems.

\bibliographystyle{ieeetr}
\bibliography{paper}

@inproceedings{mahony2017geometric,
  title={A geometric nonlinear observer for simultaneous localisation and mapping},
  author={Mahony, Robert and Hamel, Tarek},
  booktitle={2017 IEEE 56th Annual Conference on Decision and Control (CDC)},
  pages={2408--2415},
  year={2017},
  organization={IEEE}
}

@article{van2023eqvio,
  title={Eqvio: An equivariant filter for visual-inertial odometry},
  author={van Goor, Pieter and Mahony, Robert},
  journal={IEEE Transactions on Robotics},
  volume={39},
  number={5},
  pages={3567--3585},
  year={2023},
  publisher={IEEE}
}

@inproceedings{van2021autonomous,
  title={Autonomous error and constructive observer design for group affine systems},
  author={van Goor, Pieter and Mahony, Robert},
  booktitle={2021 60th IEEE Conference on Decision and Control (CDC)},
  pages={4730--4737},
  year={2021},
  organization={IEEE}
}

@article{van2023constructiveINS,
  title={Constructive Equivariant Observer Design for Inertial Navigation},
  author={van Goor, Pieter and Hamel, Tarek and Mahony, Robert},
  journal={IFAC-PapersOnLine},
  volume={56},
  number={2},
  pages={2494--2499},
  year={2023},
  publisher={Elsevier}
}

@article{VANGOOR2025112328,
title = {Synchronous observer design for Inertial Navigation Systems with almost-global convergence},
journal = {Automatica},
volume = {177},
pages = {112328},
year = {2025},
issn = {0005-1098},
doi = {https://doi.org/10.1016/j.automatica.2025.112328},
author = {Pieter {van Goor} and Tarek Hamel and Robert Mahony}
}

@book{2012_lee_IntroductionSmoothManifolds,
  title = {Introduction to {{Smooth Manifolds}}},
  author = {Lee, John M},
  date = {2012-08-27},
  year = {2012},
  series = {Graduate {{Texts}} in {{Mathematics}}},
  edition = {2},
  publisher = {Springer},
  location = {New York},
  isbn = {978-1-4419-9982-5}
}

@article{martinelli2013observability,
  title={Observability properties and deterministic algorithms in visual-inertial structure from motion},
  author={Martinelli, Agostino and others},
  journal={Foundations and Trends{\textregistered} in Robotics},
  volume={3},
  number={3},
  pages={139--209},
  year={2013},
  publisher={Now Publishers, Inc.}
}

@article{slotine1991nonlinear,
  title={Nonlinear applied control},
  author={Slotine, Jean-Jacques E and Li, Weiping},
  journal={Li, W., Ed},
  year={1991}
}

@article{vasconcelos2010nonlinear,
  title={A nonlinear position and attitude observer on SE (3) using landmark measurements},
  author={Vasconcelos, Jos{\'e} Fernandes and Cunha, Rita and Silvestre, Carlos and Oliveira, Paulo},
  journal={Systems \& Control Letters},
  volume={59},
  number={3-4},
  pages={155--166},
  year={2010},
  publisher={Elsevier}
}

@article{bailey2006slam:part1,
  title={Simultaneous localization and mapping (SLAM): Part II},
  author={Bailey, Tim and Durrant-Whyte, Hugh},
  journal={IEEE robotics \& automation magazine},
  volume={13},
  number={3},
  pages={108--117},
  year={2006},
  publisher={IEEE}
}

@article{bonnabel2008symmetry,
  title={Symmetry-preserving observers},
  author={Bonnabel, Silvere and Martin, Philippe and Rouchon, Pierre},
  journal={IEEE Transactions on Automatic Control},
  volume={53},
  number={11},
  pages={2514--2526},
  year={2008},
  publisher={IEEE}
}

@article{mahony2008nonlinear,
  title={Nonlinear complementary filters on the special orthogonal group},
  author={Mahony, Robert and Hamel, Tarek and Pflimlin, Jean-Michel},
  journal={IEEE Transactions on automatic control},
  volume={53},
  number={5},
  pages={1203--1218},
  year={2008},
  publisher={IEEE}
}

@inproceedings{baldwin2009nonlinear,
  title={A nonlinear observer for 6 DOF pose estimation from inertial and bearing measurements},
  author={Baldwin, Grant and Mahony, Robert and Trumpf, Jochen},
  booktitle={2009 IEEE International Conference on Robotics and Automation},
  pages={2237--2242},
  year={2009},
  organization={IEEE}
}

@article{barrau2015ekf,
  title={An EKF-SLAM algorithm with consistency properties},
  author={Barrau, Axel and Bonnabel, Silvere},
  journal={arXiv preprint arXiv:1510.06263},
  year={2015}
}

@article{zlotnik2018gradient,
  title={Gradient-based observer for simultaneous localization and mapping},
  author={Zlotnik, David Evan and Forbes, James Richard},
  journal={IEEE Transactions on Automatic Control},
  volume={63},
  number={12},
  pages={4338--4344},
  year={2018},
  publisher={IEEE}
}

@inproceedings{wang2018geometric,
  title={Geometric nonlinear observer design for slam on a matrix lie group},
  author={Wang, Miaomiao and Tayebi, Abdelhamid},
  booktitle={2018 IEEE Conference on Decision and Control (CDC)},
  pages={1488--1493},
  year={2018},
  organization={IEEE}
}

@article{boughellaba2025nonlinear,
  title={Nonlinear Observer Design for Landmark-Inertial Simultaneous Localization and Mapping},
  author={Boughellaba, Mouaad and Berkane, Soulaimane and Tayebi, Abdelhamid},
  journal={arXiv preprint arXiv:2504.04239},
  year={2025}
}

@inproceedings{lourencco20133,
  title={3-d inertial trajectory and map online estimation: Building on a GAS sensor-based SLAM filter},
  author={Louren{\c{c}}o, Pedro and Guerreiro, Bruno J and Batista, Pedro and Oliveira, Paulo and Silvestre, Carlos},
  booktitle={2013 European Control Conference (ECC)},
  pages={4214--4219},
  year={2013},
  organization={IEEE}
}

@inproceedings{johansen2016globally,
  title={Globally exponentially stable Kalman filtering for SLAM with AHRS},
  author={Johansen, Tor A and Brekke, Edmund},
  booktitle={2016 19th International Conference on Information Fusion (FUSION)},
  pages={909--916},
  year={2016},
  organization={IEEE}
}

@article{huang2007convergence,
  title={Convergence and consistency analysis for extended Kalman filter based SLAM},
  author={Huang, Shoudong and Dissanayake, Gamini},
  journal={IEEE Transactions on robotics},
  volume={23},
  number={5},
  pages={1036--1049},
  year={2007},
  publisher={IEEE}
}

@inproceedings{lee2006observability,
  title={On the observability and observability analysis of SLAM},
  author={Lee, Kwang Wee and Wijesoma, W Sardha and Guzman, Javier Ibanez},
  booktitle={2006 IEEE/RSJ International Conference on Intelligent Robots and Systems},
  pages={3569--3574},
  year={2006},
  organization={IEEE}
}

@article{cadena2017past,
  title={Past, present, and future of simultaneous localization and mapping: Toward the robust-perception age},
  author={Cadena, Cesar and Carlone, Luca and Carrillo, Henry and Latif, Yasir and Scaramuzza, Davide and Neira, Jos{\'e} and Reid, Ian and Leonard, John J},
  journal={IEEE Transactions on robotics},
  volume={32},
  number={6},
  pages={1309--1332},
  year={2017},
  publisher={IEEE}
}

@article{kaess2012isam2,
  title={iSAM2: Incremental smoothing and mapping using the Bayes tree},
  author={Kaess, Michael and Johannsson, Hordur and Roberts, Richard and Ila, Viorela and Leonard, John J and Dellaert, Frank},
  journal={The International Journal of Robotics Research},
  volume={31},
  number={2},
  pages={216--235},
  year={2012},
  publisher={Sage Publications Sage UK: London, England}
}

@article{JoshiBundle,
title = {A bundle framework for observer design on smooth manifolds with symmetry},
journal = {Journal of Geometric Mechanics},
volume = {13},
number = {2},
pages = {247-271},
year = {2021},
issn = {1941-4889},
doi = {10.3934/jgm.2021015},
author = {Joshi, Anant A. and Maithripala, D. H. S. and Banavar, Ravi N.}
}

@article{van2025synchronous,
  title={Synchronous Models and Fundamental Systems in Observer Design},
  author={van Goor, Pieter and Mahony, Robert},
  journal={arXiv preprint arXiv:2505.19517},
  year={2025}
}

\end{document}